\documentclass[fleqn,usenatbib]{mnras}

\usepackage{newtxtext,newtxmath}
\usepackage[T1]{fontenc}

\DeclareRobustCommand{\VAN}[3]{#2}
\let\VANthebibliography\thebibliography
\def\thebibliography{\DeclareRobustCommand{\VAN}[3]{##3}\VANthebibliography}

\usepackage{graphicx}	% Including figure files
\usepackage{amsmath}	% Advanced maths commands
\usepackage{bm}
\usepackage{orcidlink}
\title[Be star decretion disc size]{Decretion disc size in Be/X-ray binaries depends upon the disc aspect ratio  }

\author[R. G. Martin et al.]{
Rebecca G. Martin$^{1,2}$\thanks{E-mail: rebecca.martin@unlv.edu}\orcidlink{0000-0003-2401-7168},
Stephen H. Lubow$^3$\orcidlink{0000-0002-4636-7348},
Philip J. Armitage$^{4,5}$\orcidlink{0000-0001-5032-1396}
and
Daniel J. Price$^6$\orcidlink{0000-0002-4716-4235}
\\
$^1$Nevada Center for Astrophysics, University of Nevada, Las Vegas,
4505 South Maryland Parkway, Las Vegas, NV 89154, USA\\
$^2$Department of Physics and Astronomy, University of Nevada, Las Vegas,
4505 South Maryland Parkway, Las Vegas, NV 89154, USA\\
$^3$Space Telescope Science Institute, 3700 San Martin Drive, Baltimore, MD 21218, USA \\
$^4$Center for Computational Astrophysics, Flatiron Institute, 162 Fifth Avenue, New York, NY 10010, USA \\
$^5$Department of Physics and Astronomy, Stony Brook University, Stony Brook, NY 11794, USA \\
$^6$School of Physics and Astronomy, Monash University, Vic. 3800, Australia
}

\date{Accepted XXX. Received YYY; in original form ZZZ}

\pubyear{2023}

\begin{document}
\label{firstpage}
\pagerange{\pageref{firstpage}--\pageref{lastpage}}
\maketitle

\begin{abstract}
With three-dimensional hydrodynamical simulations we show that the size of the decretion disc and the structure of the accretion flow onto the neutron star in a Be/X-ray binary strongly depends upon the disc aspect ratio, $H/R$. We simulate a Be star disc that is coplanar to the orbit of a circularly or moderately eccentric neutron star companion, thereby maximising the effects of tidal truncation. For low disc aspect ratio, $H/R\lesssim 0.1$, the disc is efficiently tidally truncated by the neutron star. Most material that escapes the Roche lobe  of the Be star is accreted by the neutron star through tidal streams. For larger disc aspect ratio,  the outflow rate through the Be star disc is higher, tidal truncation becomes inefficient, the disc fills the Roche lobe and extends to the orbit of the companion. Some material escapes the binary as a gas stream that begins near the L2 point. While the accretion rate onto the neutron star is higher, the fraction of the outflow that is accreted by the neutron star is smaller. Low density Be star discs are expected to be approximately isothermal, 
such that $H/R$ increases with radius. Tidal truncation is therefore weaker for larger separation binaries, and lower mass primaries.
\end{abstract} 


%%%%%%%%%%%%%%%%%%%%%%%%%%%%%%%%%%%%%%%%%%%%%%%%%%

%%%%%%%%%%%%%%%%% BODY OF PAPER %%%%%%%%%%%%%%%%%%

\section{Introduction}

A Be/X-ray binary typically consists of a Be star with a neutron star companion \citep[e.g.][]{Negueruela1998,Coe2005,Liu2005,Reig2011,haberl2016}. The Be star is rapidly rotating \citep{Slettebak1982,Porter1996} and material that is ejected from the Be star equator forms a decretion disc \citep{Pringle1991,Lee1991,Carciofi2011}. 
The outer radial extent of a Be star decretion disc may be limited by the tidal forces from the neutron star. If the disc is geometrically thin, then the outer edge of the disc occurs at the radius where tidal and viscous forces balance \citep{Papaloizou1977,Artymowicz1994,Okazaki2002,Martin2011}. However, for circumplanetary discs with moderately large disc aspect ratio it has recently been shown that the effects of tidal truncation may be weak \citep{Martin2023}. Therefore the disc aspect ratio is an important parameter that can determine the size of disc.  

Radiative equilibrium calculations for Be star discs suggest that heating by photoionization and collisional excitation is balanced by cooling from recombination and collisional de-excitation \citep[e.g.][]{Sigut2009}. The predicted outer disc temperature is approximately isothermal, with $T \simeq 1-1.6 \times 10^4 \ {\rm K}$, and the aspect ratio $H/R$ increases with radius \citep[e.g.][]{Carciofi2006,Carciofi2006b,Rubio2023,Suffak2023}. Observationally, the disc aspect ratio for Be star discs can be measured or constrained in several different ways \citep{Porter2003,Rivinius2013}. By combining interferometry with polarimetry, an upper limit for $\zeta$-Tau was found to be $H/R=0.36$ (opening half angle of $20^\circ$) \citep{Quirrenbach1997}. Polarization can measure the inner parts of the disc, and $\zeta$-Tau was found to have an aspect ratio of $H/R=0.04$ (half-opening angle of $2.5^\circ$) \citep{Wood1997,Carciofi2009}. Be star discs are thought to be flared, meaning that $H/R$ increases with radius \citep{Hanuschik1996}. In this case, systems with a larger binary orbital period may have a larger disc aspect ratio in the outer parts of the disc. Assuming a random distribution of the Be star disc inclination on the sky, statistics of shell stars suggest that the aspect ratio could be $H/R=0.09$ (opening half-angle of $5^\circ$) \citep{Porter1996} or $H/R=0.23$ (half-opening angle of $13^\circ$) \citep{Hanuschik1996}. In general, these discs could have  aspect ratios at least in the outer parts of the disc that are large enough to lead to a weakened tidal torque. 
 
While observing the outer radius of a Be star disc is currently challenging \citep[e.g.][]{Rivinius2013}, understanding the sizes of these discs has significant implications for observations of X-ray outbursts that occur when material is transferred to the neutron star. Be/X-ray binary star systems display two different types of outburst when material falls onto the neutron star. Type~I outbursts occur each binary orbital period while Type~II outbursts are brighter and occur less frequently \citep{Stella1986,Negueruela1998,Negueruela2001,Okazaki2001,Moritani2013,Okazaki2013}. Type~I outbursts may be driven by either an eccentric binary orbit, or an eccentric disc \citep{Franchini2019b}. Type~II outbursts may be driven by a highly misaligned disc that undergoes ZKL \citep{vonZeipel1910,Kozai1962,Lidov1962} oscillations \citep{Martinetal2014,Martinetal2014b,Fu2015}.  

Tidal truncation of Be star discs has been explored extensively in theoretical works \citep[e.g.][]{Okazaki2001,Okazaki2002,Panoglou2016,Cyr2017,Suffak2022}, however, the effect of a large disc aspect ratio has not been examined. In this work we determine the disc aspect ratio required for tidal truncation of Be star disc in a binary with a neutron star. We consider an idealized model of a Be star with a spin vector that is parallel to the binary angular momentum vector such that the Be star disc is coplanar to the binary orbit. This coplanar configuration sets an upper limit to the strength of the tidal torque \citep{Lubow2015,Miranda2015}. The disc is constantly fed  gas at a particular injection radius until a steady state (or quasi-steady state in the case of an eccentric orbit binary) disc  is reached around the Be star. 
Section~\ref{1D} describes the parameters for the Be/X-ray binary. Section~\ref{hydro} presents the results of 3D simulations of the Be star disc flow in which we show that the disc aspect ratio plays a key role in determining the amount of tidal truncation and the size of a decretion disc in a Be/X-ray binary. Section~\ref{concs} contains the conclusions.

\section{Be/X-ray binary parameters}
\label{1D}

We model a Be/X-ray binary with parameters motivated by those of 4U~0115+63 \citep[e.g.][]{Campana1996,Negueruela1997}. The  Be star has a mass of $M_1=18\,\rm M_\odot$ and a radius of $R_1=8\,\rm R_\odot$. The companion neutron star has mass $M_2=1.4\,\rm M_\odot$ and is in an orbit with semi-major axis of $a_{\rm b}=95\,\rm R_\odot$. For this binary mass ratio, the Roche lobe radius of the Be star is $0.6\,a_{\rm b}=56.8{\,\rm R_\odot}$ \citep{Eggleton1983}. The binary orbit is either circular ($e_{\rm b}=0$) or mildly eccentric  ($e_{\rm b}=0.34$). For the eccentric orbit simulations, the binary begins at apastron. The disc around the Be star  is coplanar to the binary orbit so that it can reach a steady state disc structure. If the disc was inclined to the binary orbit, then it must undergo nodal precession \citep[e.g.][]{ PT1995, Bateetal2000,Lubow2000} and finding a steady state is more complex. 
  The tidal torque on a misaligned disc decreases with tilt above the orbital plane \citep{Lubow2015,Miranda2015}. Therefore the tidal truncation radius found in this work is a lower limit. A misaligned disc may be larger and have a more significant outflow of material \citep{Cyr2017}.

\begin{table*}
\begin{center}
\begin{tabular}{l c c c c c c c c c  c c l}
\hline
 Name & $H/R$ & $\alpha_{\rm AV}$ & $e_{\rm b}$  & $M_1$ & $M_2$ &$M_{\rm part}/M_{\rm s}$   & $N_{\rm steady}$ & $\left< h\right>/H$ & $\alpha$ & $\dot M_{\rm 1}/\dot M_{\rm inj}$ & $\dot M_{\rm 2}/\dot M_{\rm inj}$ & $\dot M_{\rm ej}/\dot M_{\rm inj}$  \\
 \hline
\hline
sim1 &  0.05  & 1.1 & 0.0 & 18 & 1.4&$1\times 10^{-12}$  & 76,000 & 0.74 & 0.08 & 0.978 & 0.019 & 0.003  \\  
sim2 &  0.1  & 2.9 & 0.0 & 18 & 1.4&$1\times 10^{-13}$  & 200,000 & 0.36 & 0.10 & 0.959 & 0.038 & 0.003\\
sim3 &   0.2 &  3.3 & 0.0 & 18& 1.4 &$1\times 10^{-13}$  & 95,000 & 0.30 & 0.10 & 0.932 & 0.048 & 0.019 \\
\hline
sim4 &  0.1  & 0.29 & 0.0& 18& 1.4  &$2\times 10^{-13}$  &  175,000 & 0.37 &  0.010 & 0.986 & 0.014 & 0.000  \\
sim5 &  0.2  &  0.33 & 0.0& 18 & 1.4 & $1\times 10^{-13}$  & 115,000 & 0.28 & 0.009 & 0.972 & 0.019 & 0.009\\
\hline
sim6 & 0.1  &2.9 & 0.34 & 18& 1.4 & $1\times 10^{-13}$ & 164,000 & 0.39 & 0.11 & 0.950 & 0.048 & 0.002  \\
sim7  &  0.2 &  3.3 & 0.34 & 18& 1.4 &$1\times 10^{-13}$  & 92,000 & 0.31 & 0.10 & 0.931 & 0.048 & 0.021   \\
\hline
sim8 & 0.1  & 2.9 & 0.0 &  10 & 1.4 & $1\times 10^{-13}$  & 266,000 & 0.32 & 0.09 & 0.966 & 0.034 & 0.000 \\
\hline
\end{tabular}
\end{center}
 \caption{The first seven columns describe the simulation input parameters: (left to right) the simulation name: the constant disc aspect ratio: the SPH artificial viscosity parameter $\alpha_{\rm AV}$: the binary eccentricity: the mass of the Be star: the mass of the neutron star and the mass of the SPH particles.  The next three columns describe the Be star disc the end of the simulation: (left to right) the number of SPH particles in  $R<a_{\rm b}$ (when the binary is at apastron for the eccentric orbit binaries): the density weighted average smoothing length in the disc: the calculated value for the average viscosity parameter in the steady disc. The final three columns show the binary orbital period averaged flow rates of (left to right) the accretion rate onto the Be star: the accretion rate onto the neutron star and, finally, the rate of ejection of material. 
}
\label{table}
\end{table*}

Initially there is no gas in the simulation. Material is injected into the disc close to the Be star at a constant rate of $\dot M_{\rm inj}=10^{-10}\,\rm M_\odot \, yr^{-1}$ at a radius of $R_{\rm inj}=10\,\rm R_\odot$. The injection radius is chosen to be  far from the stellar surface in order to increase the resolution of the simulations. If the injection radius is very close to the star, then an unphysically large fraction of the material falls immediately back onto the Be star \citep{Nixon2020}. This assumption changes the structure of the inner parts of the disc \citep{Rimulo2018}. However, in this work, we focus on the tidal truncation and therefore we are interested in the structure of the outer parts of the disc. 

In the absence of the tidal field, it is possible to find steady state disc solutions assuming that the viscous torque is zero at both the inner edge of the disc, $R=R_{\rm in}$, and the outer edge of the disc, $R=R_{\rm out}$. In this case, the accretion rate onto the star is given by
\begin{equation}
   \frac{ \dot M_{\rm 1}}{\dot M_{\rm inj}}=\frac{\sqrt{R_{\rm out}}-\sqrt{R_{\rm inj}}}{\sqrt{R_{\rm out}}-\sqrt{R_{\rm in}}}
    \label{analytic}
\end{equation}
\citep{Martin2023}.
Note that this estimate assumes Keplerian rotation in the disc and does not include any dependence on the disc aspect ratio, $H/R$. The decretion outflow rate is given by 
\begin{equation}
    \frac{\dot M_{\rm dec}}{\dot M_{\rm inj} }=\left(1-\frac{\dot M_1}{\dot M_{\rm inj}}\right).
\end{equation}
As the material leaves the Roche lobe of the Be star, it is  accreted onto the neutron star at a rate $\dot M_2$ and  ejected from the binary at a rate $\dot M_{\rm ej}$. Therefore we can write
\begin{equation}
\dot M_{\rm dec}=\dot M_2+\dot M_{\rm ej}.
\end{equation}
However, in the analytical model we cannot distinguish between these   two rates on the right hand side.

For our typical parameters  with a circular orbit binary, we take $R_{\rm in}=8\,\rm R_\odot$, and $R_{\rm out}=60\,\rm R_\odot$ (this assumes that the disc is tidally truncated at this radius). For these parameters with equation~(\ref{analytic}) we have $\dot M_1/\dot M_{\rm inj}=0.93$ for $R_{\rm inj}=10\,\rm R_\odot$.
Note that these estimates do not take into account the disc aspect ratio.

\begin{figure}
    \centering
    \includegraphics[width=\columnwidth]{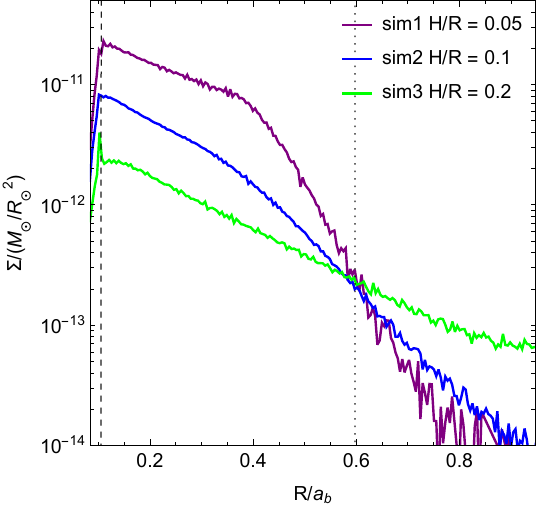}
    \caption{Azimuthally averaged steady state disc surface density profile at time $t=50\,P_{\rm orb}$ for simulations with a circular orbit binary with $M_1=18\,\rm M_\odot$, $M_2=1.4\,\rm M_\odot$ and semi-major axis of $a_{\rm b}=95\,R_\odot$. The disc has viscosity parameter $\alpha\approx 0.1$ and different disc aspect ratios, $H/R=0.05$ (sim1), $H/R=0.1$ (sim2) and $H/R=0.2$ (sim3). The vertical dashed line shows the injection radius. The vertical dotted line shows the Roche lobe radius.  }
    \label{fig:surfdens}
\end{figure}

\begin{figure*}
          \includegraphics[width=0.65\columnwidth]{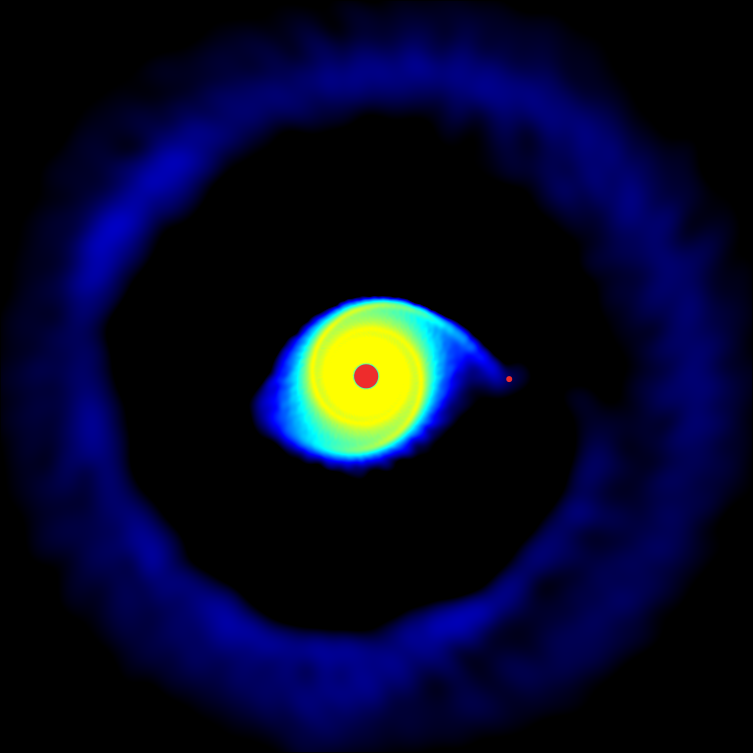}
           \includegraphics[width=0.65\columnwidth]{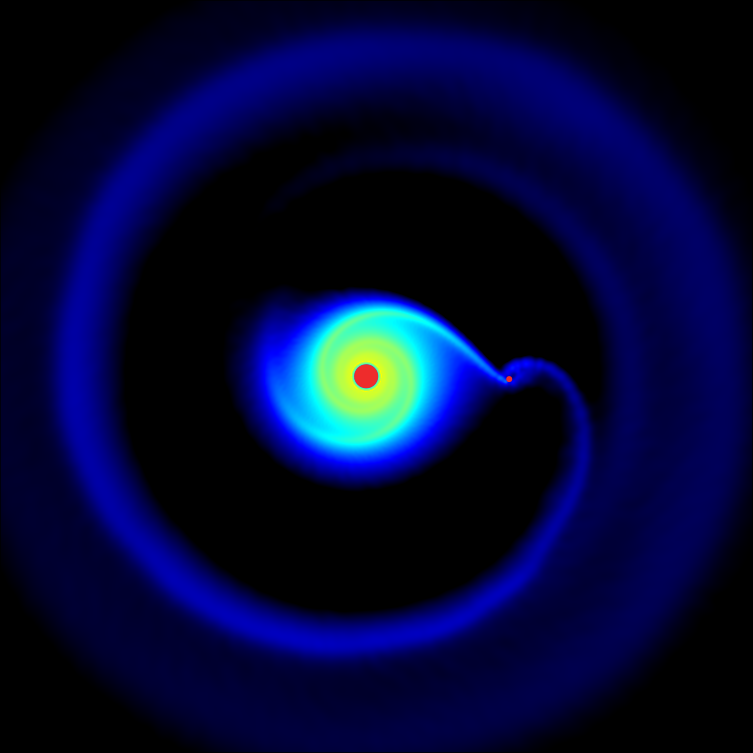}
                     \includegraphics[width=0.65\columnwidth]{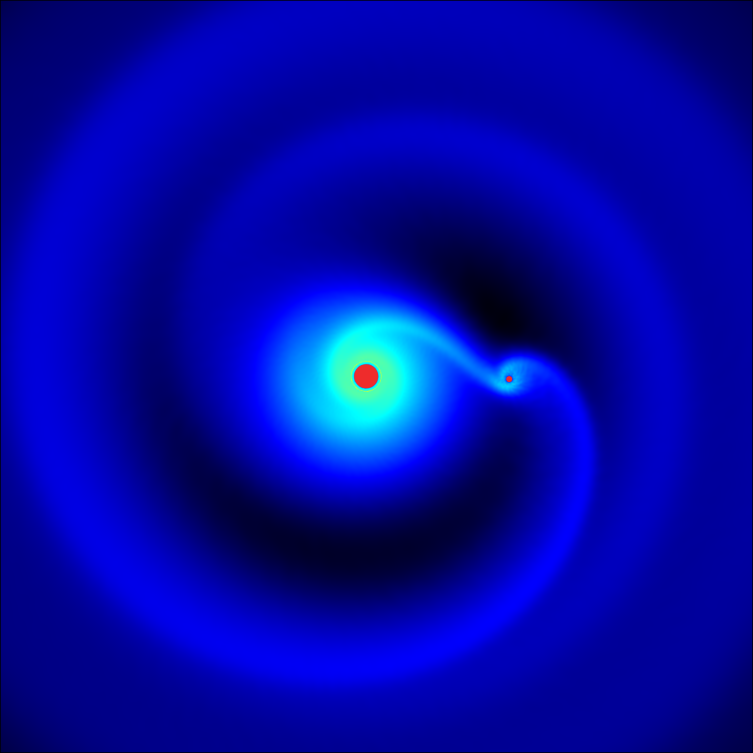}
                     \hfill
            \includegraphics[width=0.65\columnwidth]{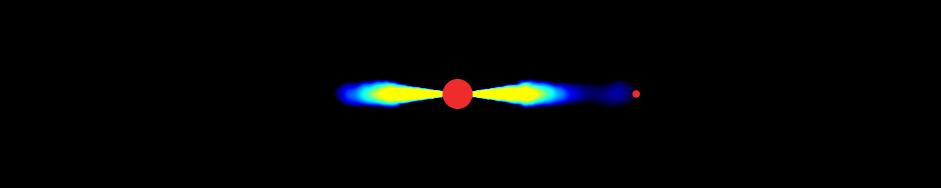}
                 \includegraphics[width=0.65\columnwidth]{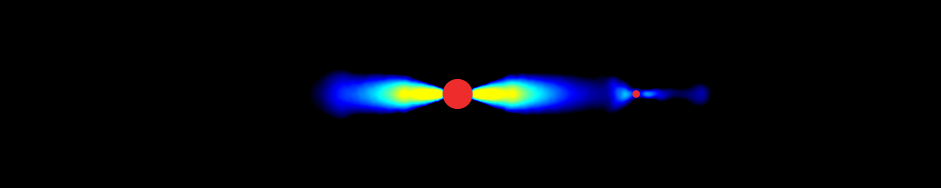}
        \includegraphics[width=0.65\columnwidth]{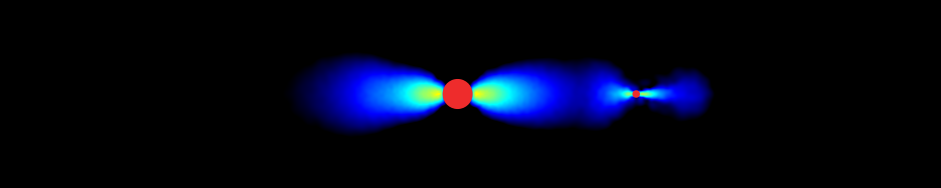}
    \caption{Visualizations of the steady state Be star disc for varying disc aspect ratio at a time of $t=50\,P_{\rm orb}$. The Be star and neutron star are shown as red solid circles with the size scaled to their accretion radius. In each panel, the upper plot shows the $x-y$ plane in which the binary orbits while the lower panel shows a cross section of the disc in the $x-z$ plane. The disc aspect ratio is  $H/R=0.05$ (left, sim1), 0.1 (middle, sim2) and 0.2 (right, sim3). The panels have a size of $500{\,\rm R_\odot} \times 500{\,\rm R_\odot}$ (upper) and  $500{\,\rm R_\odot} \times 100{\,\rm R_\odot}$ (lower).   }
    \label{fig:splash}
\end{figure*}

\section{SPH Simulations}
\label{hydro}

\begin{figure}
    \centering
     \includegraphics[width=\columnwidth]{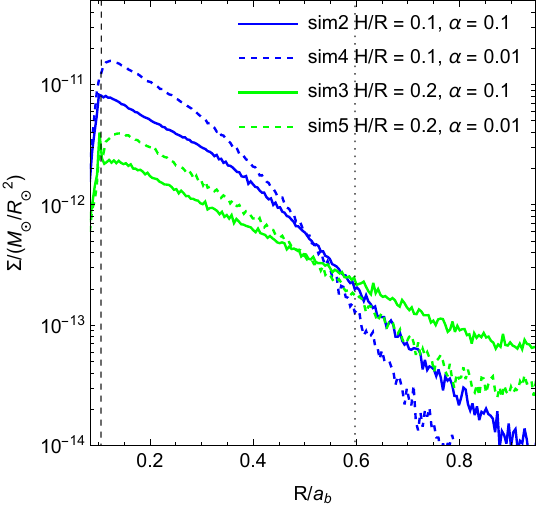}
    \caption{Same as Fig.~\ref{fig:surfdens} except the viscosity $\alpha$ parameter  is varied in the dashed lines (sim4 and sim5). The solid lines are the same as in Fig.~\ref{fig:surfdens} (sim2 and sim3).}
    \label{fig:sigma-alpha}
\end{figure}

\begin{figure}
    \centering
    \includegraphics[width=\columnwidth]{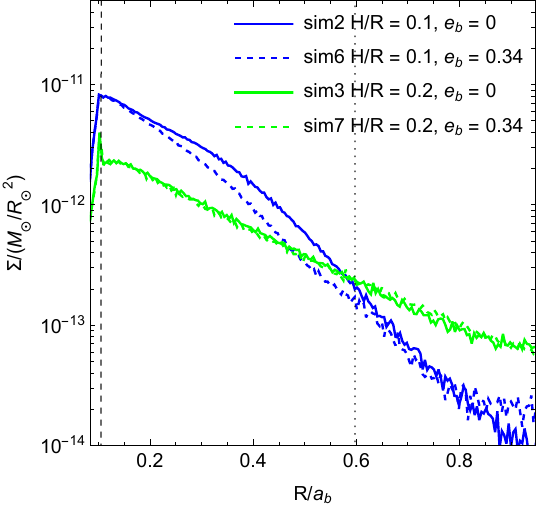}
    \caption{Same as Fig.~\ref{fig:surfdens} except the binary eccentricity  is varied in the dashed lines (sim6 and sim7). The solid lines are the same as  in Fig.~\ref{fig:surfdens} (sim2 and sim3). The surface density is shown while the binary is at apastron at time $t=50\,P_{\rm orb}$.}
    \label{fig:sigma-ecc}
\end{figure}

In order to include the effects of a tidal field from the companion neutron star and the effects of pressure, we must perform 3D hydrodynamical simulations. We use the smoothed particle hydrodynamics \citep[SPH][]{Monaghan1992,Price2012a} code {\sc phantom} \citep{Lodato2010,Price2010,Price2018}. This has been well tested for tidal truncation of discs in binary star systems \citep[e.g.][]{Franchini2019b,Franchini2019inner,Heath2020,Hirsh2020}. We note that our simulations are scaled to the injection accretion rate since we do not include disc self-gravity. The accretion rates in the steady state all scale with the mass injection rate.

The disc has a constant disc aspect ratio, $H/R$, with radius so that we can isolate its effects. The disc is locally isothermal with $T\propto R^{-1}$. The viscosity is implemented by modifying the SPH artificial viscosity. The values for $\alpha_{\rm AV}$ are shown in Table~\ref{table} and $\beta_{\rm AV}=2$. The values for $\alpha_{\rm AV}$ are chosen so that the \cite{SS1973} viscosity parameter, $\alpha$, is close to either 0.1 or 0.01. This value is calculated with equation~(38) in \cite{Lodato2010}.  It is not easy to obtain a predetermined value of $\alpha$
in the simulations because it depends upon the smoothing length $\left<h\right>$ that changes with the surface density in the steady state disc. This requires iterating on values for $\alpha_{\rm AV}$ in several simulations. Be stars are likely fully ionised and therefore have a relatively large viscosity, of the order of 0.1 \citep{Jones2008,Carciofi2012,Ghoreyshi2018,Rimulo2018,Martin2019,Granada2021}. We consider a lower value in addition for comparison.

The simulations are evolved for a time of $t= 50\,P_{\rm orb}$, where $P_{\rm orb}$ is the  binary orbital period, except for sim4 that is evolved for $100\,P_{\rm orb}$. This is sufficient time for the Be star disc to reach a steady state.  In the steady state, the number of particles in $R<a_{\rm b}$, $N_{\rm steady}$, averaged over an orbital period is constant in time. Table~\ref{table} shows the input parameters of the simulations (the first 7 columns) and the properties of the steady state disc (next 3 columns). The smoothing length, $\left<h\right>/H$, and the \cite{SS1973} viscosity $\alpha$ parameter are density weighted averages over the disc. The final three columns show the accretion rate onto the Be star, onto the neutron star and the ejection rate in the steady state averaged over an orbital period. 

\subsection{Effect of the disc aspect ratio}

The first three simulations in Table~\ref{table} show the effect of increasing the disc aspect ratio. 
Fig.~\ref{fig:surfdens} shows the surface density profile for the Be star discs at the end of the simulations. The larger the disc aspect ratio the larger the size of the Be star disc.
Fig.~\ref{fig:splash} shows the column density of the system viewed from face on and an edge on cross section for the same simulations. 
There is a transition in the size of the disc for $H/R\approx 0.1$.  For smaller disc aspect ratio, the disc is effectively tidally truncated. However, for larger disc aspect ratio the tidal truncation becomes weaker. Disc material overflows the Roche lobe of the Be star and extends to the neutron star where a small disc forms around the neutron star \citep[e.g.][]{Hayasaki2004,Hayasaki2006}. Material can also flow from the L2 point of the neutron star as a gas stream and this forms circumbinary material \citep[e.g.][]{Shu1979}.

The final three columns in Table~\ref{table} show that for larger disc aspect ratio, less material is accreted onto the Be star and more material flows outwards. The accretion rate onto the neutron star and the ejection rate both increase with disc aspect ratio.  This is a result of a number of factors. First, a larger disc aspect ratio leads to a weaker tidal torque  and a stronger viscous torque. This allows more material to flow outside of the Roche lobe of the Be star. Second, and a much smaller effect, is that the material is being added at Keplerian velocity while the velocity of the disc becomes more sub-Keplerian \citep[see Section~3.7 in][]{Martin2023}.
The ratio of the material accreted onto the neutron star compared to the material ejected decreases with increasing disc aspect ratio.

\subsection{Effect of the viscosity parameter}

We consider two simulations with a value for the $\alpha$ viscosity parameter that is smaller by a factor of 10, sim4 and sim5. Comparing these with sim2 and sim3 shows that if the value for $\alpha$ is decreased by a factor of 10, then the truncation efficiency, as defined by the accretion rate onto the star, is not significantly affected.    Fig.~\ref{fig:sigma-alpha} shows the steady state surface density profile in comparison to the larger $\alpha$ simulations. As expected, the simulations with a smaller $\alpha$ parameter have less material at larger radii since the truncation  is a little more efficient. However, the relatively small change to the truncation efficiency for a large change to the viscosity parameter suggests that the truncation is more sensitive to the disc aspect ratio than the $\alpha$ parameter.

\subsection{Effect of the binary eccentricity}

We also consider some simulations with a non-zero binary eccentricity of $e_{\rm b}=0.34$ in sim6 ($H/R=0.1$) and sim7 ($H/R=0.2$). In general, the mild binary eccentricity does not significantly change the orbital period averaged accretion rates. The surface density profile for these is shown at apastron binary separation in Fig.~\ref{fig:sigma-ecc} along with the circular orbit simulations for comparison.  For the larger disc aspect ratio simulation ($H/R=0.2$) that is not tidally truncated, there is little difference in the surface density profile with the increased eccentricity. There is also little change to the accretion rates onto the stars and the ejection rate.  For the smaller disc aspect ratio simulation that is tidally truncated ($H/R=0.1$), there is some difference in the surface density profile. This is because the tidal truncation radius of a disc scales with the binary eccentricity like $(1-e_{\rm b})$ \citep{Artymowicz1994, Hirsh2020}. The smaller disc size leads to slightly larger outflow compared to accretion onto the Be star.

\subsection{Effect of the binary mass ratio, $q=M_2/M_1$}

Finally we consider a simulation with a smaller mass for the Be star, sim8, that has binary mass ratio of $q=M_2/M_1=0.14$. Comparing this with sim2 that has $q=0.078$ shows that the larger mass ratio leads to a more efficient tidal truncation. In this case, there is no significant mass ejection from the binary. The accretion rate onto the neutron star is slightly lower compared with sim2.

\section{Conclusions}
\label{concs}

We have performed hydrodynamic simulations of the formation of a steady state Be star decretion disc to examine its interaction with a companion neutron star. We consider the case that the Be star disc is coplanar to the binary orbital plane therefore maximising the tidal truncation efficiency.
We have shown that the disc aspect ratio plays a crucial role in the truncation efficiency. The larger the disc aspect ratio, the larger the rate of outflow through the disc and the larger the size of the disc. A disc with a small disc aspect ratio $H/R\lesssim 0.1$,  is tidally truncated. The outflowing material is mostly accreted onto the neutron star through tidal streams. For larger disc aspect ratio,  tidal truncation is weak and the disc the fills the Roche lobe and extends out to the orbit of the neutron star. In this case, a large fraction of the outflowing material is ejected from the binary system.  The effects of binary eccentricity are negligible on the disc structure for discs with large aspect ratio.

\section*{Acknowledgements}
Computer support was provided by UNLV’s National Supercomputing Center. RGM and SHL acknowledge support from NASA through grants 80NSSC21K0395 and 80NSSC19K0443. PJA and RGM acknowledge support from NASA TCAN award 80NSSC19K0639. 
We acknowledge the use of SPLASH \citep{Price2007} for the rendering of Fig.~\ref{fig:splash}. The paper arose from our mutual interaction at the Kavli Institute for Theoretical Physics program on ``Bridging the Gap: Accretion and Orbital Evolution in Stellar and Black Hole Binaries", funded by the National Science Foundation under Grant No. NSF PHY-1748958. SHL thanks the Simons Foundation for support of a visit to the Flatiron Institute.

%%%%%%%%%%%%%%%%%%%%%%%%%%%%%%%%%%%%%%%%%%%%%%%%%%
\section*{Data Availability}

 The data underlying this article will be shared on reasonable request to the corresponding author.
 
%The inclusion of a Data Availability Statement is a requirement for articles published in MNRAS. Data Availability Statements provide a standardised format for readers to understand the availability of data underlying the research results described in the article. The statement may refer to original data generated in the course of the study or to third-party data analysed in the article. The statement should describe and provide means of access, where possible, by linking to the data or providing the required accession numbers for the relevant databases or DOIs.



\bibliographystyle{mnras}
\bibliography{mainmnras} % if your bibtex file is called example.bib

%\begin{thebibliography}{99}
%\end{thebibliography}

%%%%%%%%%%%%%%%%%%%%%%%%%%%%%%%%%%%%%%%%%%%%%%%%%%

%%%%%%%%%%%%%%%%% APPENDICES %%%%%%%%%%%%%%%%%%%%%

%\appendix



%%%%%%%%%%%%%%%%%%%%%%%%%%%%%%%%%%%%%%%%%%%%%%%%%%

% Don't change these lines
\bsp	% typesetting comment
\label{lastpage}
\end{document}